\documentstyle[aps,multicol,prl,tighten,graphicx]{revtex}
\begin{document}
\draft
\title{Quantum Trajectories for Realistic Detection}
\author{P. Warszawski${}^{1}$, H.M. Wiseman${}^{1,2}$, and H.
Mabuchi${}^{2}$}
\date{\today}
\address{${}^{1}$ School of Science, Griffith University, Nathan, Brisbane,
Queensland 4111 Australia\\
${}^{2}$ Norman Bridge Laboratory of Physics 12-33, California
Institute of Technology, Pasadena, CA 91125, USA}
\maketitle

\begin{abstract}
Quantum trajectories describe the stochastic evolution of an open quantum
system conditioned on continuous monitoring of its output, such as by an
ideal photodetector. Here we derive (non-Markovian) quantum trajectories for
{\em realistic} photodetection, including the effects of efficiency, dead
time, bandwidth, electronic noise, and dark counts. We apply our theory to a
realistic cavity QED scenario and investigate the impact of such detector
imperfections on the conditional evolution of the system state. A practical
theory of quantum trajectories with realistic detection will be essential for
experimental and technological applications of quantum feedback in
many areas.
\end{abstract}

\newcommand{\beq}{\begin{equation}}
\newcommand{\eeq}{\end{equation}}
\newcommand{\bqa}{\begin{eqnarray}}
\newcommand{\eqa}{\end{eqnarray}}
\newcommand{\nn}{\nonumber}
\newcommand{\nl}[1]{\nn \\ && {#1}\,}
\newcommand{\erf}[1]{Eq.~(\ref{#1})}
\newcommand{\erfs}[2]{Eqs.~(\ref{#1})--(\ref{#2})}
\newcommand{\dg}{^\dagger}
\newcommand{\rt}[1]{\sqrt{#1}\,}
\newcommand{\smallfrac}[2]{\mbox{$\frac{#1}{#2}$}}
\newcommand{\half}{\smallfrac{1}{2}}
\newcommand{\bra}[1]{\langle{#1}|}
\newcommand{\ket}[1]{|{#1}\rangle}
\newcommand{\ip}[2]{\langle{#1}|{#2}\rangle}
\newcommand{\sch}{Schr\"odinger }
\newcommand{\schs}{Schr\"odinger's }
\newcommand{\hei}{Heisenberg }
\newcommand{\heis}{Heisenberg's }
\newcommand{\bl}{{\bigl(}}
\newcommand{\br}{{\bigr)}}
\newcommand{\ito}{It\^o }
\newcommand{\str}{Stratonovich }
\newcommand{\dbd}[1]{\frac{\partial}{\partial {#1}}}
\newcommand{\sq}[1]{\left[ {#1} \right]}
\newcommand{\cu}[1]{\left\{ {#1} \right\}}
\newcommand{\ro}[1]{\left( {#1} \right)}
\newcommand{\an}[1]{\left\langle{#1}\right\rangle}
\newcommand{\implies}{\Longrightarrow}
\newcommand{\ve}{\varepsilon}

\begin{multicols}{2}

\section{Introduction}
The limited utility of quantum measurement theory as axiomatized by von
Neumann \cite{Von32} for describing practical laboratory measurements has
necessitated the development of more general measurement theories
\cite{Dav76,Kra83}. In the past decade the application of such theories has
become widespread in quantum optics, in particular for describing continuous
monitoring of the photoemission from radiatively damped open systems. They
describe the evolution of the conditioned system state in terms of quantum
jumps \cite{DalCasMol92,GarParZol92,Car93b} for direct detection and quantum
diffusion \cite{Car93b,WisMil93c} for dyne detection. The
stochastic evolution equation,  termed a quantum trajectory, has also
been applied in mesoscopic electronics \cite{WisWah01}.

Thus far, the main practical utility of quantum trajectory theory has been in
improving the computational efficiency of simulations used to compare models
with experimental data. But it is now gaining increasing importance as the
quantum generalization of Kalman filtering, which provides essential
signal-processing methods in classical estimation, communication, and control
engineering. Quantum trajectory theory should in principle play the same
pivotal role for emerging quantum analogs of these technologies
\cite{Gamb01,Cira97,Dohe99b}. Before this can happen it is essential that the
theory be extended to account for the imperfections of {\em realistic
measurement devices}, as non-ideal detector dynamics can dramatically affect
the proper inference from measured signals to the conditional quantum state
of an observed system.

In this paper we present the theory of quantum trajectories for realistic
photodetection. We model both photon counters and photoreceivers (for
homodyne detection) and include the effects of efficiency, dead time,
bandwidth, electronic noise, and dark counts. The proper treatment of
bandwidth limitations and electronic noise are of particular significance as
these imperfections are inevitable and predominant concerns in any practical
context. They are of central importance in the current generation of
experiments on quantum-limited measurement in atomic \cite{MabYeKim99} and
condensed matter \cite{Dev00} systems.

Our theory works by embedding the system within a supersystem that obeys a
Markovian equation. If the set of (classical) detector states is ${\bf S}$,
then the supersystem is described by the set $\{\rho_{s}:s \in {\bf S}\}$.
Here ${\rm Tr}[\rho_{s}]$ is the probability that the apparatus is in state
$s$, and $\rho_{s}/{\rm Tr}[\rho_{s}]$ is the system state given this event.

\section{The System}  
In this paper we take the
 monitored system to be a two-level atom (TLA), classically driven at
 Rabi frequency $\Omega$ and radiatively
damped at rate $\Gamma$.
The TLA obeys the unconditional master equation (ME)
 \beq
\dot{\rho}={\cal L}\rho=-i(\Omega/2)\left[\sigma_{x},
\rho\right]+\Gamma
\left(\sigma\rho\sigma^{\dag}-\half\{\sigma^{\dag}\sigma,\rho\}\right),
\label{TLAME} \eeq
where  $\sigma$ is the
atomic lowering operator, and $\sigma_{x}=\sigma+\sigma\dg$. Time
arguments are not included unless they are necessary for
 the reader's understanding.  In reality, it is difficult to detect a
 significant fraction of an atom's fluorescence. However, the ME
 (\ref{TLAME}) also describes, in a suitable regime \cite{Rice88},
 the damping of an atom through a cavity mode. This produces
 an easily detectable output beam. In this scenario, the effective decay rate
 $\Gamma$ may be much larger than that of a bare atom, and we have
 this in mind when choosing $\Gamma=300$M$s^{-1}$ for our simulations.

\section{Photon Counter}  
\begin{figure}
\includegraphics[width=0.45\textwidth]{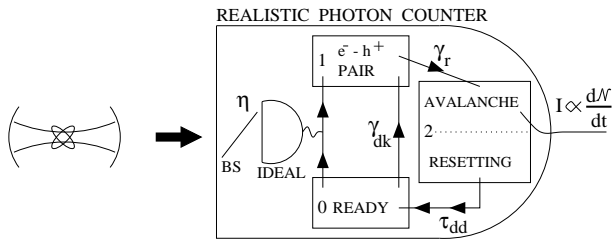}
\vspace{0.2cm}
\caption{\narrowtext  Realistic photon counting by an avalanche
photodiode.
The quantum efficiency
$\eta$ is represented by the beam splitter (BS).
 Single arrowheads within the realistic photodetector
 indicate Poisson processes. For details, see text.}
\protect\label{PDDiag}
\end{figure}

An avalanche photodiode (APD) operating in
Geiger mode produces a
macroscopic current pulse in response
to an incident photon.  It consists
of a p-n junction
 operated under a reverse bias greater than the breakdown voltage
\cite{OpFibV1}. Under these
conditions we can describe the diode by just three classical
states (see Fig.~\ref{PDDiag}). The first ($0$) is a stable low-current state
in which
there are no charge carriers in the depletion region of the
junction.
 The transition from $0$ to the second state ($1$) takes place when
  an electron--hole
pair is created in the depletion region
by an incident photon (with quantum efficiency $\eta$) or by
thermally initiated
`dark counts' occurring at a rate $\gamma_{{\rm dk}}$.
Further
impact ionization leads to an avalanche, until the current
reaches some threshold value and a detection is registered, thus changing
the state of the APD to $2$.  The transition from $1$ to $2$
has a random duration (we assume Poissonian)
with mean $\gamma_{\rm r}^{-1}$ (the `response time').
The avalanche is then arrested by the application of a
negative-going voltage pulse that temporarily brings the bias voltage below the
breakdown value \cite{OpFibTech}.  This results in a fixed
 `dead time', $\tau_{{\rm dd}}$, during which the APD cannot
detect photons, after which it is restored to state 0.

Our aim is to derive the quantum trajectories for the quantum system (the
source of the light entering the APD) {\em conditioned on the observation of
an avalanche}. For the TLA, we consider direct detection of fluorescence. In
this case the supersystem is described by the set
($\tilde{\rho}_{0},\tilde{\rho}_{1},\tilde{\rho}_{2})$, where the tilde
indicates that (for simplicity) we are using unnormalized system states, and
the subscript indicates the associated detector states. The normalized
conditioned TLA state is \beq \rho_{\rm c} = \tilde{\rho}_{\rm c}/{\rm
Tr}[\tilde{\rho}_{\rm c}] \;;\;\; \tilde{\rho}_{{\rm c}}=
\tilde{\rho}_{0}+\tilde{\rho}_{1}+\tilde{\rho}_{2}. \eeq
Our description of internal dynamics of an APD can be simply translated into
rate equations for the discrete detector state $s\in\{0,1,2\}$, which in turn
imply the following stochastic generalization of the ME~(\ref{TLAME}):
\bqa
d\tilde{\rho}_{0}&=&dt\left\{\left[{\cal L}-\gamma_{{\rm
dk}}- \eta\Gamma{\cal
J}-\dot{{\cal N}}\right]\tilde{\rho}_{0}+{\dot
{\cal N}}(t^{*}) \tilde{\rho}_{2}\right\}, \label{dp0}\\
d\tilde{\rho}_{1}&=&dt\left\{\left[{\cal L}-\gamma_{{\rm
r}}-\dot{{\cal N}}\right]\tilde{\rho}_{1}+\eta\Gamma{\cal J}\tilde{\rho}_{0}
+\gamma_{{\rm dk}}\tilde{\rho}_{0}\right\}\label{dp1},\\
d\tilde{\rho}_{2}&=&dt\left\{\left[{\cal
L}-\dot{{\cal N}}(t^{*})\right]\tilde{\rho}_{2}+\dot{{\cal
N}}\tilde{\rho}_{1}\right\}.
\label{dp2}
\eqa
Here ${\cal
J}\tilde\rho_{0}$ denotes $\sigma\tilde\rho_{0}\sigma\dg$. We use ${\cal
N}$ for the
number of detections counted, so that $d{\cal N}(t)=\dot{{\cal N}}dt$ is a
point process
equal to $1$ in the infinitesimal interval when an avalanche is first
observed and $0$ otherwise. The delayed process, $\dot{{\cal N}}(t^{*})\equiv
\dot{{\cal N}}(t-\tau_{{\rm dd}})$, is used to return the detector to state
$0$. The
statistics of $d{\cal N}$ are defined by its expectation value $ {\rm
E}[d{\cal N}]=\gamma_{{\rm r}}dt{\rm Tr}[\tilde{\rho}_{1}]/{\rm
Tr}[\tilde{\rho}_{\rm c}]$.

\begin{figure}
\includegraphics[width=0.45\textwidth]{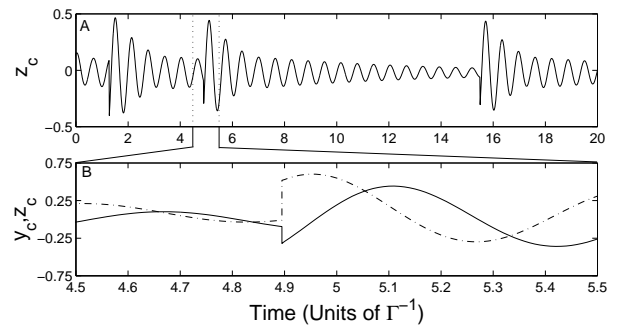}
\vspace{0.2cm}
\caption{\narrowtext In plot (A),
$z_{{\rm c}}={\rm Tr}[(2\sigma\dg\sigma-1)\rho_{\rm c}]$ is shown for
a typical trajectory of duration $20\Gamma^{-1}$.  In plot (B), $z_{{\rm
c}}$ and $y_{{\rm c}}=-i{\rm Tr}[(\sigma-\sigma\dg)\rho_{\rm c}]$
(dash-dot) are
shown for the same trajectory near the time of the second APD
avalanche. The realistic
parameters used for this photodetection trajectory were
$\eta=80\%$, and (in units of the TLA decay
rate $\Gamma=300$M$s^{-1}$) $\gamma_{{\rm r}}=7$,
$\tau_{{\rm dd}}=2$,
$\gamma_{{\rm dk}}=5\times 10^{-6}$ and $\Omega=10$.}
\protect\label{TrajsPRL}
\end{figure}

The detector imperfections lead to substantial changes in the conditional
dynamics of the TLA, as compared to ideal quantum trajectories.
Representative features  can be seen in
Fig.~\ref{TrajsPRL} (A) and (B). Plot (A) shows a typical portion of a
trajectory for
$z_{{\rm c}}$, while plot (B) shows the same, and  $y_{{\rm c}}$,
over a shorter time around $t\approx 4.9$, when an avalanche is
registered. Unlike the case of ideal detection, the
corresponding ``quantum jump'' does not take $z_{{\rm c}}\rightarrow
-1,y_{{\rm c}}\rightarrow 0$, and the amplitude of subsequent oscillations in
$z_{{\rm c}},y_{{\rm c}}$ are less than $1$. The jumps in the conditioned
quantum state caused by the detection of avalanches are attenuated because of
the finite detector response time in combination with the continuous Rabi
oscillation, which evolves the TLA away from the ground state for a random
and unknown time (with mean $\gamma_{{\rm r}}^{-1}$) between the ``actual''
spontaneous emission event and the registration of the photocurrent avalanche.
During the APD dead time the effective efficiency is
zero, and as a result the TLA's conditional state regresses towards the
steady state of the unconditional ME (1). Even after the detector becomes
ready again (by resetting to state 0) the Rabi oscillations in $z_{\rm
c},y_{\rm c}$ decay because of the APD's non-unit efficiency and
finite bandwidth.

These imperfections cause the stationary
ensemble-averaged conditional
purity $p = \lim_{t\to\infty}{\rm E}\cu{{\rm Tr}[\rho_{\rm
c}^{2}(t)]}$ to be substantially
less than one for large $\Omega$.
For small $\Omega$ however, even the unconditional (without measurement)
stationary purity $p_{\rm u}$ of the TLA approaches unity. It is thus useful
to define a scaled purity $\in [0,1]$ that measures how much
improvement measurement gives: ${\rm Scaled\;}p=(p-p_{{\rm u}})/(1-p_{{\rm
u}}).$
For the typical parameter values used in Fig.~\ref{TrajsPRL},
the Scaled $p \approx 0.052$.

\section{Photoreceiver}
\begin{figure}
\includegraphics[width=0.45\textwidth]{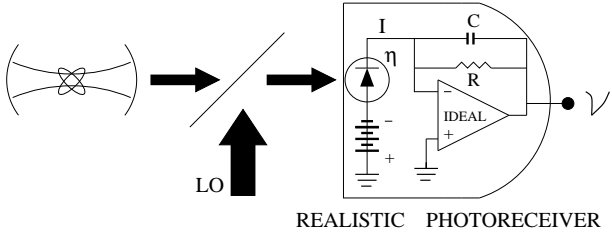}
\vspace{0.2cm} \caption{\narrowtext  Homodyne detection by a
photoreceiver. The output field of the TLA is combined with a LO
 before being detected by a realistic
photoreceiver consisting of a p-i-n photodiode (of
quantum efficiency $\eta$)
that produces the photocurrent $I$, and an ideal op-amp with
feedback resistor $R$, and capacitance, $C$ (see text).
The output voltage ${\cal V}$ is what is measured in the
laboratory.} \protect\label{PRDiag}
\end{figure}

  When the incident photon flux is high, as in
homodyne detection, a p-i-n photodiode connected to a transimpedance
amplifier (see Fig.~\ref{PRDiag}) is an appropriate photoreceiver
\cite{OpFibTech}. When a photon strikes the depletion region of the p-i-n
junction, an electron--hole pair
 is produced, with probability equal to the quantum
efficiency $\eta$. The charge carriers drift under the influence of the
below-breakdown reverse bias, and the resultant
 current $I$ is fed into an operational amplifier (op-amp) set up as a
 transimpedance amplifier. This has a low effective input impedance,
 so that the diode acts as a current source, and $I$ is converted into
  a voltage drop $V$ across the
feedback resistor, $R$.  The
capacitor $C$, in parallel with $R$, represents the total
capacitance from the output of the op-amp back to its input,
including capacitance added deliberately for the smoothing of noise
and oscillations.  If no electronic noise were present, the output
voltage of the photoreceiver would be a filtered version of the input
signal given, in the frequency domain, by
\beq
V(\omega )={-IR}/({1+i\omega RC}).
\label{Vfreq1}
\eeq

It should be noted that if this were the case (that is, if there were no
noise) then the input $I$ could be perfectly reconstructed from the filtered
signal $V$. Thus the resultant quantum trajectories would be no different
from those of a photoreceiver with infinite bandwidth. Everything of interest
results therefore from the presence of excess noise. We include only the
Johnson noise $V_{{\rm J}}$ from the feedback resistor, which has a flat
spectrum $S_{{\rm J}}=4k_{{\rm B}}TR$. This simplification (neglecting
contributions from voltage noise of the operational amplifier) can be
justified for practical receivers with $R\sim 10$k$\Omega$.

The output voltage ${\cal V}$ from the photoreceiver is given by the sum
 of the filtered signal and the Johnson noise
\beq {\cal V}=V+V_{{\rm J}}.
\label{Vo}
\eeq
Our aim is to find the quantum trajectory for the system, conditioned
on continuously monitoring ${\cal V}$. Since the voltage $V$, which
describes the detector state,
is a continuous variable, in this case ${\bf S}= {\rm I\!R}$, the real line,
and the supersystem can be described by an operator function
$\rho(V)$. Finding the stochastic equation of motion for $\rho(V)$ is quite
involved.

We begin by taking the output current $I$ of the photodiode to be that from a
perfect (apart from its efficiency $\eta$) unbalanced
homodyne detection
of the fluorescence of the TLA.
For a LO tuned to the atomic transition frequency
$\omega_{0}$,  of power $P$, and
phase $\phi$, the current is \cite{Car93b,WisMil93a}
 \beq
 I=e\sqrt{P/\hbar\omega_{0}}\left[\eta\sqrt{\Gamma}\langle e^{-i\phi}\sigma
+e^{i\phi}\sigma^{\dag}\rangle +\rt{\eta}\xi(t)\right],
\eeq
where we have ignored the D.C. component due to the LO power. Here
$\xi(t)$ is the Gaussian white noise \cite{Gar85}
 arising from the Poissonian statistics of the LO and
 $e$ is the electron charge.
The evolution of the TLA conditioned on $I$ is given, in terms of
the noise $\xi(t)$, by the
following stochastic master equation \cite{WisMil93a}
\beq
d\rho_{I}=dt\left\{{\cal
L}+\rt{\eta\Gamma}\xi(t){\cal H}[e^{-i\phi}\sigma]\right\}\rho ,
 \label{rhoI}
 \eeq
 where ${\cal H}[A]\rho\equiv A\rho+\rho
A^{\dag}-{\rm Tr}[A\rho+\rho A^{\dag}]\rho$.

Now \erf{Vfreq1} is equivalent to the
stochastic  equation
\beq
I+V/R+C(dV/dt)=0.
\label{LE}
\eeq
Since the voltage $V$ is not directly measured, we must consider a
distribution $P(V)$ for it. Assuming that $C>0$, and, for the moment,
 that $I$ is known,  \erf{LE} can be converted to
an \ito \cite{Gar85} stochastic Fokker-Planck equation
for $P(V)$ conditioned on the photocurrent:
\beq dP_{I}(V)=\left(\frac{\partial}{\partial
V}\frac{V+IR} {RC}+\frac{P\eta
e^{2}}{2\hbar\omega_{0}C^{2}}\frac{\partial^{2}} {\partial
V^{2}}\right)P(V)dt.
\label{PI}
\eeq
Here we are using the convention that subscripts
indicate that the increment is conditioned
on that result.  That is, for example, $P_{I}(V)\equiv P(V|I)$.

Next we need to determine the effect of the measurement of ${\cal V}$
on $P(V)$.  This can be
calculated by using Bayes' conditional probability theorem
\beq
P_{{\cal V}}(V)=P_{V}({\cal V})P(V)/P({\cal V}).
\label{bayes}
\eeq
Remembering that the
Johnson noise is white,
 it follows from \erf{Vo} that
$P_{V}({\cal V})$ is a Gaussian with mean $V$ and variance $4k_{{\rm
B}}TR/dt$. From this we find that
\beq
P({\cal V})=\int dVP_{V}({\cal V})P(V).
\eeq
 is a Gaussian of mean
$\langle V\rangle$ and
variance $4k_{{\rm B}}TR/dt$.  It follows that we can write
\beq \label{defdWJ}
{\cal V}=\langle V\rangle+\rt{4k_{{\rm B}}TR}{dW_{{\rm J}}(t)}/{dt},
\eeq
where $dW_{{\rm J}}(t)/dt$ is another Gaussian white noise source,
independent of $\xi(t)$.
Substitution of $P_{V}({\cal V})$ and $P({\cal V})$ into \erf{bayes}
yields the effect of the ${\cal V}$-measurement:
\beq
dP_{{\cal V}}(V)=
\left(V-\langle V\rangle\right)P(V)dW_{{\rm J}}(t)/\sqrt{4k_{{\rm B}}TR}.
\label{PV}
\eeq

Now, to see how ${\cal V}$ conditions the TLA, we form the
quantity $\rho(V)=\rho P(V)$,
where $\rho$ is here independent of $P(V)$ because we are imagining
$I$ to be known at all times.
The time evolution of $\rho(V)$, given that ${\cal V}$ and $I$ are
known, is found from
\bqa
\rho(V)+d\rho_{I,{\cal V}}(V)&=&\left(\rho+d\rho_{I}\right)\nl{\times}
\left[P(V)+dP_{I}(V)+dP_{{\cal V}}(V)\right],
\eqa
with the use of \erf{PI},
\erf{PV} and \erf{rhoI}.

Finally, in reality, ${\cal V}$ is known but $I$ is
not. Therefore we should average over the vacuum noise $\xi(t)$,
but keep the Johnson noise $dW_{\rm J}/dt$. We define a dimensionless voltage
$v=V\sqrt{C/4k_{{\rm B}}T}$, a rate $\gamma=1/RC$ and a dimensionless noise
power
$N = 4k_{{\rm B}}T\hbar\omega_{0}/\eta RPe^{2}$. This last expression is
the ratio of the
low-frequency power in ${\cal V}$ from the
Johnson noise to that from the vacuum noise. We then obtain the
following stochastic nonlinear superoperator Fokker-Planck equation for
$\rho(v)$:
\bqa
d\rho_{{\cal V}}(v)&=&dt\left({\cal
L}+\frac{\gamma}{2N}\frac{\partial^{2}}{\partial
v^{2}}+\gamma\frac{\partial}{\partial v}v \right)\rho(v)
\nl{+}dt\frac{\partial}{\partial
v}\sqrt{\frac{\gamma\Gamma\eta}{N}}
\left[e^{-i\phi}\sigma\rho(v)+
e^{i\phi}\rho(v)\sigma^{\dag}\right]
\nl{+}
\rt{\gamma}dW_{{\rm J}}(t)
\left(v-\langle v\rangle\right)\rho(v).
\label{dpHom}
\eqa
 The dependence on ${\cal V}$
 may be explicated by substituting
$dt\gamma\ro{\rt{{C}/{ 4k_{\rm B}T}} {\cal V}-\an{v}}$
for
$\rt{\gamma}dW_{{\rm J}}(t)$ [see \erf{defdWJ}]. In the above,  $\langle
v\rangle=\int dv{\rm Tr}[\rho(v)]v$. The
normalized conditioned TLA state is  $\rho_{\rm
c}=\int\rho(v)dv$.

\begin{figure}
{\bf See attached file 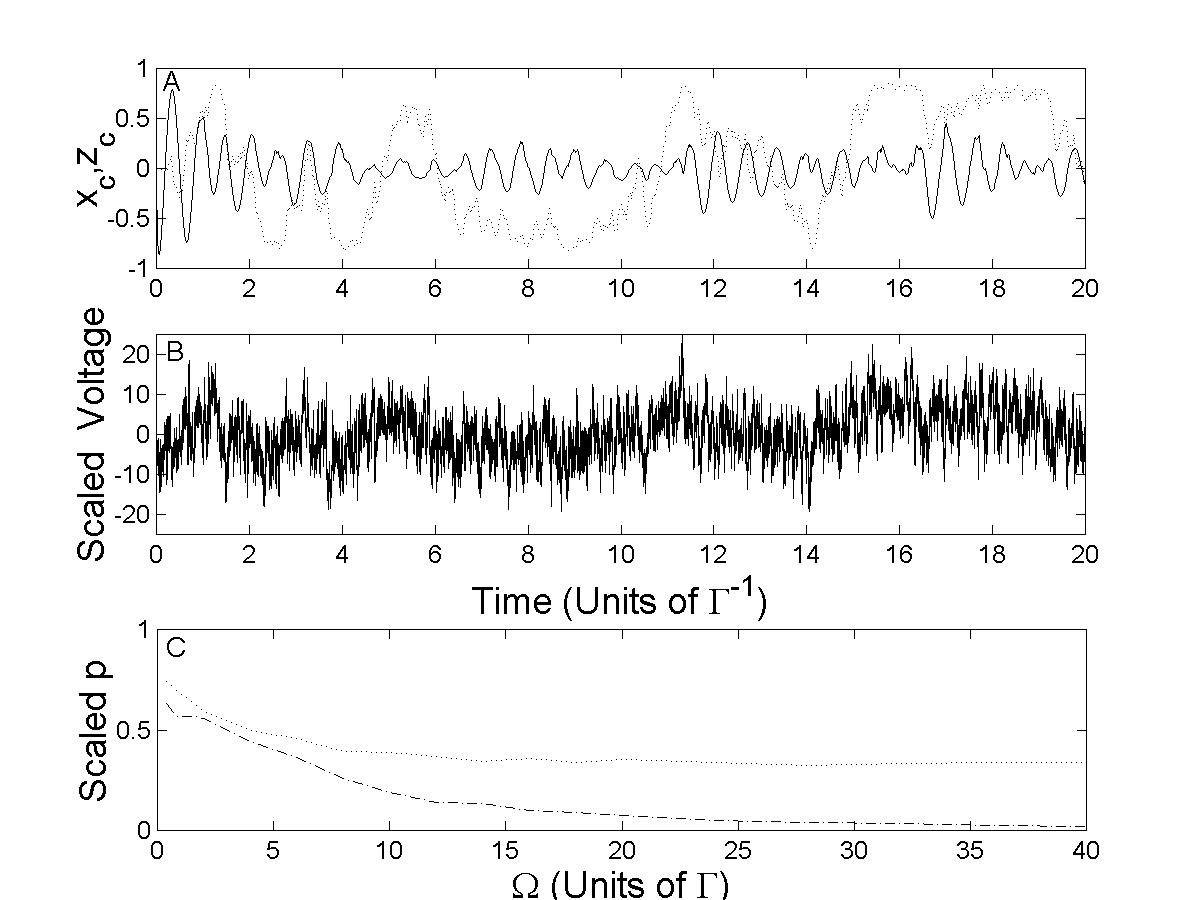}
\vspace{0.2cm}
\caption{\narrowtext Plot (A) shows
$x_{{\rm c}}={\rm Tr}[(\sigma+\sigma\dg)\rho_{\rm c}]$ (dotted) and
$z_{{\rm c}}$ (solid) for a realistic homodyne $x$ trajectory. Plot (B)
is the dimensionless output voltage from the photoreceiver.
The photoreceiver parameters were
$N=0.1$,    $\eta=98\%$, $\gamma=1.5\Gamma$. System parameters were as
for Fig.~\ref{TrajsPRL}.  Plot C
gives the Scaled $p$ as a function of the driving, $\Omega$, for
homodyne $x$ (dotted) and $y$ (dash-dot) detection.}
\protect\label{TrajsPRLHom}
\end{figure}

 A Typical trajectory for realistic homodyne $x$ ($\phi=0$)
 detection is shown in
Fig.~\ref{TrajsPRLHom} (A).  The main difference from the case
of perfect detection is the reduced amplitude of variation in
 $x_{{\rm c}}$ (dotted) and  $z_{{\rm c}}$ (solid).
 This is due to the effective bandwidth of the photoreceiver,
 which affects $z_{\rm c}$ more because of its faster dynamics.
Plot (B) shows the photoreceiver
output voltage ${\cal V}$ that is used in
\erf{dpHom} to condition the TLA state. It is seen that
${\cal V}$ is correlated with $x_{\rm c}$ as expected.

Plotted in (C) is the scaled purity as a
function of the driving strength for both homodyne $x$ and $y$
($\phi=\pi/2$) detection. As $\Omega$ increases, homodyne $y$
detection becomes increasingly worse than $x$ detection at following the
evolution of the TLA.  This is due to the finite bandwidth of the
photoreceiver in combination with the conditional homodyne dynamics in the
$\Omega \gg \Gamma$ limit \cite{WisMil93c}.  For homodyne $x$ measurement the
$x$ quadrature, which changes sign fairly infrequently, dominates the TLA
state.  The slow ($\Gamma$) dynamics allow the detector to track of the
state reasonably well.  In
contrast, homodyne $y$ detection produces a conditional state
dominated by fast ($\Omega$) Rabi cycling, which is poorly followed.

\section{Conclusions}

In conclusion, we have presented a theory of quantum trajectories for systems
conditioned on realistic photodetection. The equations are tractable, as we
have demonstrated by numerical simulations, and allow us to quantify the
degree and manner by which imperfections such as a finite bandwidth modify the
conditioning of quantum states by measurement in concrete experimental
scenarios. Realistic quantum trajectory models will be of paramount
importance in the field of real-time quantum feedback control
\cite{Dohe2000}. The techniques we introduce here may also prove
essential in describing other realistic measurements, such as in condensed
matter systems
\cite{WisWah01,Dev00}.

\end{multicols}

\end{document}